\documentclass[a4paper, 10pt, conference]{ieeeconf}      

\IEEEoverridecommandlockouts                              
\usepackage[utf8]{inputenc}
\usepackage{graphicx}

\title{\LARGE \bf
Summary of the 4th International Workshop on Requirements Engineering and Testing (RET 2017)
}

\author{%
Markus Borg$^{1}$, Elizabeth Bjarnason$^{2}$, Michael Unterkalmsteiner$^{3}$, Tingting Yu$^{4}$, Gregory Gay$^{5}$, and Michael Felderer$^{6}$
\thanks{$^{1}$M. Borg is with RISE SICS AB, Sweden.
        {\tt\small markus.borg@ri.se}}%
\thanks{$^{2}$E. Bjarnason is with Lund University, Sweden
        {\tt\small elizabeth.bjarnason@cs.lth.se}}%
\thanks{$^{3}$M. Unterkalmsteiner is with Blekinge Institute of Technology, Sweden
        {\tt\small mun@bth.se}}%
\thanks{$^{4}$T. Yu is with University of Kentucky, USA
        {\tt\small tyu@cs.uky.edu}}%
\thanks{$^{5}$G. Gay is with University of South Carolina, USA
        {\tt\small greg@greggay.com}}%
\thanks{$^{6}$M. Felderer is with University of Innsbruck, Austria
        {\tt\small michael.felderer@uibk.ac.at}}
}

\begin{document}

\maketitle
\thispagestyle{empty}
\pagestyle{empty}

\begin{abstract}
The RET (Requirements Engineering and Testing) workshop series provides a meeting point for researchers and practitioners from the two separate fields of Requirements Engineering (RE) and Testing. 
The goal is to improve the connection and alignment of these two areas through an exchange of ideas, challenges, practices, experiences and results. 
The long term aim is to build a community and a body of knowledge within the intersection of RE and Testing, i.e. RET. 
The 4th workshop was co-located with the 25th International Requirements Engineering Conference (RE'17) in Lisbon, Portugal.
In line with the previous workshop instances, RET 2017 offered an interactive setting with a keynote, an invited talk, paper presentations, and a concluding hands-on exercise.
The workshop attracted about 20 participants and the positive feedback we received encourages us to organize the workshop again next year.
\end{abstract}

\section{Introduction}
The main objective of the RET workshop series is to explore, characterize, and understand the interaction of Requirements Engineering (RE) and software Testing, both in research and industry, and the challenges that result from this interaction. 
RET provides a forum for exchanging experiences, ideas, and best practices to coordinate RE and testing. 
One of the major goals of this exchange is to enable and provide incentives for research that both crosses research areas and is relevant for industry. 
RET supports this goal by inviting submissions exploring how to coordinate RE and Testing, including practices, artifacts, methods, techniques and tools. 
Furthermore, RET welcomes submissions also on softer aspects like the communication between roles in engineering processes.

RET 2017 accepted technical papers with a maximum length of 8 pages presenting research results or industrial practices and experiences related to the coordination of RET, as well as position papers with a maximum length of 4 pages introducing challenges, visions, positions or preliminary results within the scope of the workshop. 
New for this year was the call for tool papers up to 4 pages long.
As always, RET particularly welcomed experience reports and papers on open challenges in industry.

RET 2017 accepted five technical papers and three short papers, incl. a short paper with a tool focus.
The workshop attracted about 20 participants\footnote{Exact numbers are hard to provide as RE participants registered for all workshops and often moved between rooms.} and the proceedings are available online~\cite{c1}.
Finally, an image gallery from the workshop is available on the RET 2017 webpage: http://ret.cs.lth.se/17/

\section{Organization}
The 4th International Workshop on Requirements Engineering and Testing (RET 2017) was held on September 5 2017, co-located with the 25th International Requirements Engineering Conference (RE'17). 
The roles of the workshop organization follow:
\begin{itemize}
\item Markus Borg, general chair
\item Elizabeth Bjarnason, program co-chair
\item Tingting Yu, program co-chair
\item Michael Unterkalmsteiner, media chair
\item Gregory Gay, co-chair
\item Michael Felderer, co-chair
\end{itemize}

Moreover, RET has a steering committee consisting of senior researchers in requirements engineering and/or software testing:
\begin{itemize}
\item Jane Cleland-Huang, DePaul University, USA
\item Mats Heimdahl, University of Minnesota, USA
\item Jane Huffman Hayes, University of Kentucky, USA
\item Marjo Kauppinen, Aalto University, Finland
\item Per Runeson, Lund University, Sweden
\item Paolo Tonella, Fondazione Bruno Kessler, Italy
\end{itemize}

\section{Program Summary}
The program of RET 2017 comprised of four sessions: an introductory session with a keynote, two paper sessions including an invited talk, and an interactive session with a hands-on exercise.

\subsection{Session 1}
After a welcome note by Markus Borg, presenting the history and evolution of the RET workshop series, Thomas Olsson gave a keynote. Olsson has a licentiate degree in Software Engineering and has worked as a researcher with Fraunhofer, focusing on quality requirements. After the focus on research, Thomas worked almost a decade with product management at Sony Ericsson, later renamed to Sony Mobile. Olsson's talk was entitled ``Coordinating requirements engineering and quality assurance in a complex release planning context: Experience from Sony Mobile''.  

Olsson presented the overall context When planning the release of new software for a mobile phone at Sony mobile, highlighting the many factors influencing the planing -- including several limitations inhibiting an ideal process. In particular, Olsson stressed that 1) basically the same software is used globally, 2) the mobile network operators typically require at least two test rounds which need to be scheduled weeks in advance, 3) despite the best efforts, avoiding a big bang like integration is almost impossible, 4) the main drivers for lead-time is not under Sony's control, namely when Google will release the next update, and 5)  type approval is needed to be allowed to release a new software. All these factors influence how new features can be planned, agreed, implemented, and tested for various stakeholders. The release environment depicted was clearly complex, and Thomas shared his views on what could maybe have been done differently.

\subsection{Session 2}
The second session started with an invited talk on a systematic mapping study that reviewed the scientific literature addressing requirements engineering and software testing alignment. Karhap\"a\"a presented his work, previously published at the 21st Conference on Evaluation and Assessment in Software Engineering (EASE 2017)~\cite{c4}, in which the authors identified 7 focus areas in the 80 reviewed publications. Prior to the workshop, we used their structure to create a visual map of research on RET alignment, including their assessments of rigor and relevance~\cite{c3}.

\begin{figure*}
\begin{center}
\scalebox{0.75}{\includegraphics{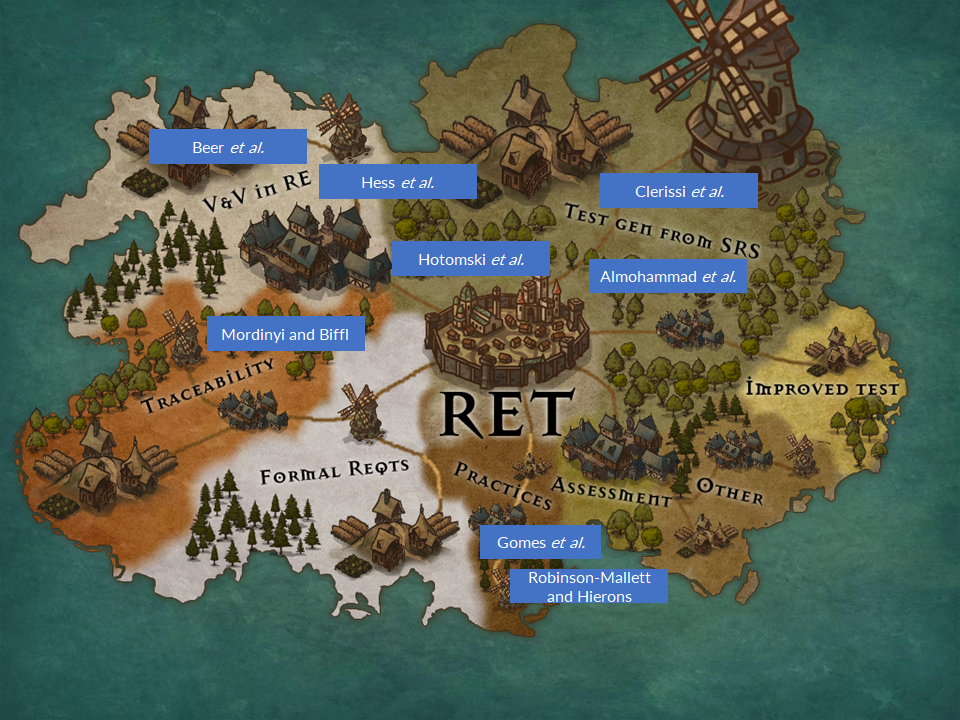}}
\caption{A visual map of research on RET alignment.}
\label{fig:map}
\end{center}
\end{figure*}

\begin{figure}
\begin{center}
\scalebox{0.3}{\includegraphics{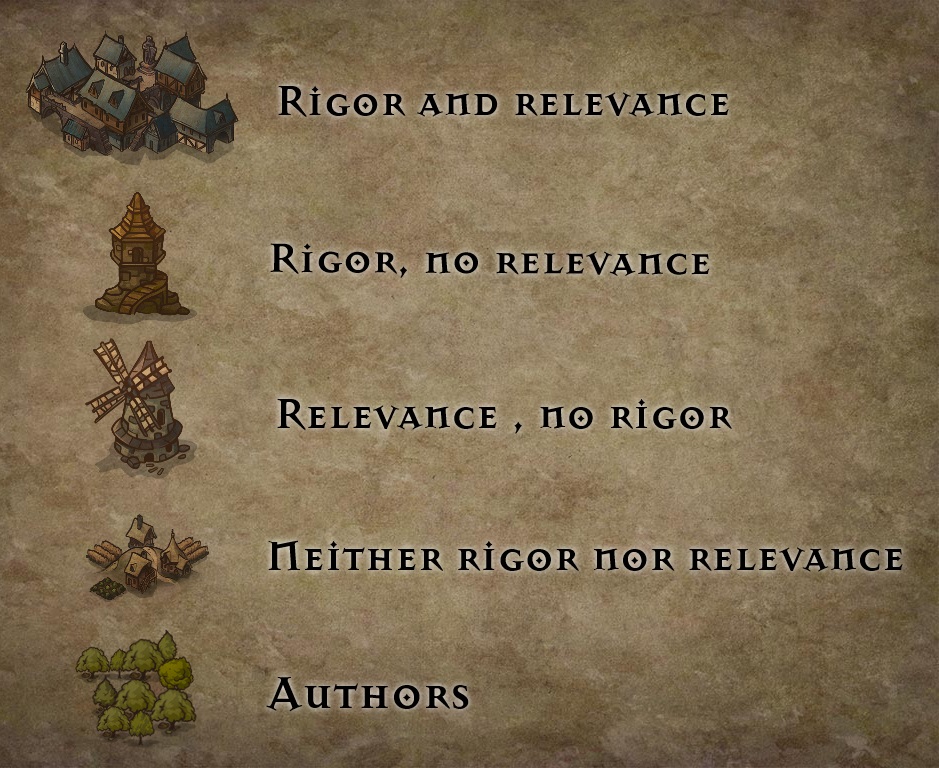}}
\caption{Legend of the visual RET map.}
\label{fig:legend}
\end{center}
\end{figure}

Figure~\ref{fig:map} shows the visual RET map, created using the Inkarnate\footnote{http://inkarnate.com} fantasy map generator. We used the map during the remainder of the sessions to categorize the papers presented at the workshop (cf. the author names in blue boxes). Figure~\ref{fig:legend} presents the legend of the map. The eight regions reflect the focus areas identified in Karhap\"a\"a \textit{et al.}'s work, their respective sizes reflect the number of corresponding previous publications. Each tree in a region displays the number of authors that have published in the focus area. Rigor and relevance of the previous work are represented using four different buildings, and their sizes reflect the number of previous publications, i.e., more publications equals bigger building. \textit{Towns} show work with both rigor and relevance, (ivory) \textit{towers} show work with rigor but little relevance, \textit{windmills} represent studies with relevance but low rigor, and farms show work with neither high rigor nor high relevance. Finally, the capital with city walls in the middle of the map represents the RET workshop -- the event that connects all regions of RET. Note that there are no ivory towers on the map, suggesting that RET is a generally relevant software engineering topic.

The session continued with the presentation of one full and two short papers. Gomes presented work resulting from a multiple case study, investigating the challenges of aligning requirements engineering and system testing in the large-scale agile context. In addition to illustrate the studied cases and describing the challenges in terms of risk and prevalence among the interviewed companies, the authors identify Agile requirements engineering practices that contain potential to address those challenges.

Femmer presented work that investigates two what extent quality problems in requirements documents influence test-case design. The authors compared the prevalence of test-case design antipatterns resulting from requirements that were seeded with faults and from requirements that did not contain those faults. This initial investigation indicates that the comprehensibility of requirements has an effect on the quality of test-cases based on those requirements.

Hess presented a tool to filter information contained in software requirements specifications (SRS) based on the information needs of a particular role. This approach can be used to reduce the complexity of SRS documents, allowing for example testers to find information efficiently. The views are priority-based and can be adapted based on the particular needs of the company.

\subsection{Session 3}
The third session comprised five paper presentations. Robinson-Mallett shared experience from the automotive domain in Germany in the talk ``Integrating graphical and natural language specifications to support analysis and testing''. The authors address the challenge of basing testing on low-quality requirements specifications. They propose combining natural language requirements with a graphical notation, i.e., model integration, both for improving existing requirements and for developing an SRS from scratch. The approach is already mature enough to have been implemented in tools used in industry. 

Hotomski presented a short paper titled ``Aligning requirements and acceptance tests via automatically generated guidance''. The overall research topic is maintaining RET alignment as artifacts keep changing during a development project. The authors note that previous work on change alerts merely reports that a change has occurred. In this work, however, the authors took the first steps toward implementing a tool that also provides guidance on how impacted artifacts should be updated. 

Leotta gave a talk entitled ``Towards the generation of end-to-end web test scripts from requirements specification'' –- the only talk this year targeting web development. Their novel solution generates test scripts from textual or UML-based requirements specifications. The rigorous specification work required to apply the approach suggests that the primary candidates for technology transfer include critical web applications, e.g., banking, insurance, and governmental crisis response management.

Almohammad presented a requirements modelling tool that generates test cases according to code coverage criteria such as decision and condition coverage. The tool has been evaluated through a case study and is used internally at the company.

Mordinyi presented work on providing a requirements coverage metrics based on assessing code coverage of related test cases.  The connection between requirements and test cases is derived from information provided by the engineers in the version control system.

\subsection{Session 4}
The workshop papers were further discussed and analyzed in smaller groups during the interactive exercise. A visual abstract for design science research was used to pinpoint and highlight the addressed problem and the proposed solution for each paper. A set of questions were provided to support a structured discussion around these aspects in smaller groups. The aim of the approach is to support communicating the research and highlighting its contributions by eliciting clear and crisp descriptions of the main aspects of the research.

The authors that participated in the exercise, found that the visual abstract for design science provided structured highlighted aspects that they had considered when preparing their presentations. They also expressed that the provided discussion questions were useful in supporting a structured discussion around the research and help to convey these aspects. However, it took some time to communicate the context of the situation. One presenter expressed that the exercise help to clarify the core problem addressed by her research, which covers several related problems and effects. The participants also expressed that pinpointing the problem, solution and technological rule became easier as they gained experience of the method and knew what to extract.

\section{Topic analysis}
\begin{figure}
\includegraphics[width=\columnwidth]{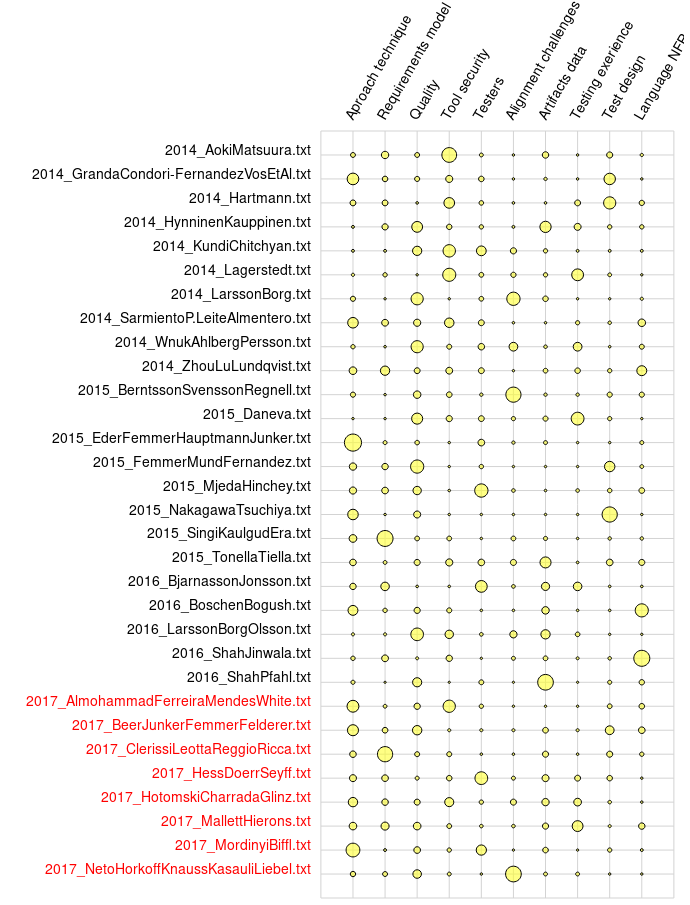}
\caption{Topic model of the papers accepted at RET 2014-2017.}
\label{fig:tm}
\end{figure}

As in the previous instance of the workshop, we created a topic model from the, now 30, abstracts of the papers that were accepted at the workshop between 2014 and 2017. We used Serendip~\cite{c2} to create and visualize the topic model (see Figure~\ref{fig:tm}). The rows represent the papers presented at RET in three years. The columns represent the identified topics\footnote{The number of topics, 10, is a required parameter when generating the model and was set rather arbitrarily. However, 10 topics seemed to be enough to provide some differentiation between papers and not too much to be too fine-grained. Most of the topics were rather easy to label, based on the most frequent terms per topic.}. The size of the circle on the crossing between article and topic represents the probability that the document was generated by the terms that represent the respective topic. The predominant topics at the respective workshop instances were:

\begin{itemize}
\item RET 2014:  Tools, Security Requirements, System Testing, Experience, Development and Models
\item RET 2015:  Test Design, Testers, Quality
\item RET 2016:  Language, Quality, Artifacts and Data
\item RET 2017:  Approach technique, Requirements model, Quality
\end{itemize}

In Figure~\ref{fig:tm}, the papers from RET 2017 are highlighted in red and the respective topics are ordered from left to right, by the total proportion. The dominant topic this year was ``Approach technique'', which suggests that (in line with RET 2014) the accepted papers were solution oriented. Furthermore, ``quality'' was again a popular topic at RET, as has been the case since 2015.

\section{Future}
The RET workshop once again well-received and attracted a mix of participants from academia, research institutes, and industry.
Since the topic remains relevant, we plan to organize the workshop again next year. Our aim is to again co-locate RET with RE, thus our target is for RET 2018 to be co-located with the 40th International Software Engineering Engineering Conference (ICSE'18) in Gothenburg, Sweden.  If the workshop is accepted, the expected date for paper submissions is in February 2018.

\section{acknowledgements}
Thanks go to the participants of the workshop and all the authors of submitted papers for their important contribution to the event. In addition, we want to thank the organizers of RE'17 and the members of the RET program committee:

\begin{itemize}
\item Armin Beer, Beer Test Consulting, Austria
\item Ruzanna Chitchyan, Leicester University, UK
\item Nelly Condori-Fernandez, VU University Amsterdam, NL
\item Robert Feldt, Chalmers University of Technology, Sweden
\item Henning Femmer, Technische Universität München, Germany
\item Joel Greenyer, University of Hannover, Germany
\item Andrea Herrmann, Herrmann \& Ehrlich, Germany
\item Mike Hinchey, Lero - the Irish Software Engineering Research Centre, Ireland
\item Jacob Larsson, Capgemini, Sweden
\item Emmanuel Letier, University College London, UK
\item Annabella Loconsole, Malm\"o University, Sweden
\item Alessandro Marchetto, Independent Researcher, Italy
\item Mirko Morandini, W\"urth Phoenix, Italy
\item Magnus C. Ohlsson, System Verification, Sweden
\item Dietmar Pfahl, University of Tartu, Estonia
\item Sanjai Rayadurgam, University of Minnesota, USA
\item Giedre Sabaliauskaite, Singapore University of Technology and Design, Singapore 
\item Kristian Sandahl, Link\"oping University, Sweden
\item Ola S\"oder, Axis Communications AB, Sweden
\item Hema Srikanth, IBM, USA
\item Marc-Florian Wendland, Fraunhofer FOKUS, Germany
\end{itemize}

\addtolength{\textheight}{-12cm}  


\end{document}